\documentclass[sigconf,authorversion]{acmart}
\usepackage[english]{babel}
\usepackage{blindtext}
\usepackage{acro}
\usepackage{siunitx}
\usepackage{xurl}

\usepackage{tabularx,booktabs}
\usepackage{multirow, makecell}
\usepackage{xfp}
\usepackage{ifthen}
\usepackage{bm}
\usepackage{nicefrac}
\usepackage{subcaption}
\usepackage[inline]{enumitem}
\usepackage{colortbl}

\usepackage{prettyref}

\acsetup{use-id-as-short}
\DeclareSIUnit{\noop}{\kern 0pt}

\newcount\tmpnum

\def\storedataA#1{\advance\tmpnum by1
\ifx\end#1\else
\expandafter\def\csname data\tmp\the\tmpnum\endcsname{#1}%
\expandafter\storedataA\fi
}

\def\assert#1{\ifthenelse{#1}{}{\errmessage{ASSERT FAIL}}}
\def\getarrdata[#1]#2{\ifcsname data#2#1\endcsname\csname data#2#1\endcsname\else\errmessage{UNSET #2#1}\fi}

\def\getarr[#1]#2{\getarrdata[#1]{#2}}

\def\roundprefixprefix[#1]#2{\SI[scientific-notation = engineering, exponent-to-prefix = true, round-mode=places,round-precision=#1]{#2}{\noop}}

\def\roundprefix[#1]#2{\ifthenelse {#2 < 1000}{#2\,\,\,~}{\roundprefixprefix[#1]{#2}}}

\def\round[#1]#2{\SI[round-mode=places,round-precision=#1,round-integer-to-decimal]{#2}{\noop}}

\def\rsienv#1{\roundprefix[2]{\fpeval{#1}}}
\def\rsienvshort#1{\roundprefix[0]{\fpeval{#1}}}

\def\rgetarr[#1]#2{\ifthenelse{\equal{\getarr[#1]{#2}}{ }}{}{\roundprefix[2]{\getarr[#1]{#2}}}}
\def\ngetarr[#1]#2{\ifthenelse{\equal{\getarr[#1]{#2}}{ }}{}{\SI[group-separator={\,}, group-minimum-digits=4]{\getarr[#1]{#2}}{\noop}}}

\def\calcperc#1#2{$\sim$\SI[round-mode=places,round-precision=0]{\fpeval{((#1)*1.0)/((#2)*1.0)*100}}{\percent}}

\def\calcpercprec[#1]#2#3{$\sim$\SI[round-mode=places,round-precision=#1]{\fpeval{((#2)*1.0)/((#3)*1.0)*100}}{\percent}}
\def\calcpercprecnosim[#1]#2#3{\SI[round-mode=places,round-precision=#1]{\fpeval{((#2)*1.0)/((#3)*1.0)*100}}{\percent}}

\newcommand{\ie}{i.e.,}
\newcommand{\eg}{e.g.,}
\newcommand{\etal}{et~al\@ifnextchar.{}{.\@}}
\newcommand{\etc}{etc\@ifnextchar.{}{.\@}}

\newcommand{\cf}{cf. }

\newcommand{\afblock}[1]{\noindent{\textbf{#1}}}
\newcommand{\takeaway}[1]{\noindent{\textbf{Takeaway.}} \textit{#1}}

\definecolor{plotblue}{rgb}{0,0.42,0.65}
\definecolor{plotred}{rgb}{0.95,0.42,0.39}

\newrefformat{chap}{Chapter~\ref{#1}}
\newrefformat{sec}{Section\,\ref{#1}}
\newrefformat{sub}{Section\,\ref{#1}}
\newrefformat{subsec}{Section\,\ref{#1}}
\newrefformat{ssub}{Section\,\ref{#1}}
\newrefformat{eq}{Equation~(\ref{#1})}
\newrefformat{fig}{Figure~\ref{#1}}
\newrefformat{plot}{Figure~\ref{#1}}
\newrefformat{ex}{Example~\ref{#1}}
\newrefformat{list}{List~\ref{#1}}
\newrefformat{tab}{Table~\ref{#1}}
\newrefformat{enum}{Case~(\ref{#1}.)}
\newrefformat{def}{Definition~\ref{#1}}
\newcommand{\pref}[1]{\prettyref{#1}}

\DeclareAcronym{ECN}{long=Explicit Congestion Notification}
\DeclareAcronym{AQM}{long=Active Queue Management}
\DeclareAcronym{ECT}{long=ECN-Capable Transport} %

\copyrightyear{2023}
\acmYear{2023}
\setcopyright{acmlicensed}
\acmConference[IMC '23]{Proceedings of the 2023 ACM Internet Measurement Conference}{October 24--26, 2023}{Montreal, QC, Canada}
\acmBooktitle{Proceedings of the 2023 ACM Internet Measurement Conference (IMC '23), October 24--26, 2023, Montreal, QC, Canada}
\acmPrice{15.00}
\acmDOI{10.1145/3618257.3624821}
\acmISBN{979-8-4007-0382-9/23/10}

\begin{document}

\title{ECN with QUIC: Challenges in the Wild}

\author{Constantin Sander}
\orcid{0009-0004-6627-1708}
\affiliation{%
\institution{RWTH Aachen University}
\city{Aachen}
\country{Germany}
}
\email{sander@comsys.rwth-aachen.de}

\author{Ike Kunze}
\orcid{0000-0001-8609-800X}
\affiliation{%
\institution{RWTH Aachen University}
\city{Aachen}
\country{Germany}
}
\email{kunze@comsys.rwth-aachen.de}

\author{Leo Blöcher}
\orcid{0009-0001-4263-6483}
\affiliation{%
\institution{RWTH Aachen University}
\city{Aachen}
\country{Germany}
}
\email{bloecher@comsys.rwth-aachen.de}

\author{Mike Kosek}
\orcid{0000-0002-3299-5546}
\affiliation{%
\institution{Technical University of Munich}
\city{Munich}
\country{Germany}
}
\email{kosek@in.tum.de}

\author{Klaus Wehrle}
\orcid{0000-0001-7252-4186}
\affiliation{%
\institution{RWTH Aachen University}
\city{Aachen}
\country{Germany}
}
\email{wehrle@comsys.rwth-aachen.de}

\renewcommand{\shortauthors}{Constantin Sander, Ike Kunze, Leo Blöcher, Mike Kosek, \& Klaus Wehrle}




\begin{abstract}
TCP and QUIC can both leverage ECN to avoid congestion loss and its retransmission overhead.
However, both protocols require support of their remote endpoints and it took two decades since the initial standardization of ECN for TCP to reach \SI{80}{\percent} ECN support and more in the wild.
In contrast, the QUIC standard mandates ECN support, but there are notable ambiguities that make it unclear if and how ECN can actually be used with QUIC on the Internet.
Hence, in this paper, we analyze ECN support with QUIC in the wild:
We conduct repeated measurements on more than 180\,M domains to identify HTTP/3 websites and analyze the underlying QUIC connections w.r.t. ECN support.
We only find \SI{20}{\percent} of QUIC hosts, providing \SI{6}{\percent} of HTTP/3 websites, to mirror client ECN codepoints.
Yet, mirroring ECN is only half of what is required for ECN with QUIC, as QUIC validates mirrored ECN codepoints to detect network impairments:
We observe that less than \SI{2}{\percent} of QUIC hosts, providing less than \SI{0.3}{\percent} of HTTP/3 websites, pass this validation. %
We identify possible root causes in content providers not supporting ECN via QUIC and network impairments hindering ECN.
We thus also characterize ECN with QUIC distributedly to traverse other paths and discuss our results w.r.t. QUIC and ECN innovations beyond QUIC.
\end{abstract}

\maketitle
\section{Introduction}
\ac{ECN}~\cite{RFC3168} allows network devices to signal incipient congestion via explicit markings instead of relying on the implicit dropping of packets.
End-to-end congestion control can thus react prior to actual packet loss, avoiding latency and bandwidth overheads of retransmissions. %
While supporting \ac{ECN} requires an interplay of the network path of a flow and ECN mirroring on the transport layer, related work finds ECN support on more than \SI{80}{\percent}~\cite{kuehlewind:tma18:ecnpathspider, lim:arxiv22:freshlookecn} of HTTP webservers reachable via TCP roughly two decades after its introduction, thereby creating the basis for successful use of ECN on the Internet.%

The modern web in the form of HTTP/3, however, switches away from TCP to QUIC~\cite{RFC9000} and is seeing an increasing adoption~\cite{rueth:pam18:gquicwild, zirngibl:imc21:quic9000}.
Nowadays, we find one fourth of the \texttt{com/net/org} domains accessible via TCP-based HTTP to also be accessible via HTTP/3 and QUIC.
In contrast to TCP, however, there is still no clear picture on ECN support with QUIC.
On the one hand, the QUIC standard mandates ECN mirroring, i.e., signaling ECN back to remote endpoints necessary to react on ECN marks, if the required network layer information can be accessed~\cite{RFC9000}.
As such, multiple open source QUIC stacks do mirror and use ECN~\cite{quicinterop}, or are in the process of adding support~\cite{googleecn1,lsquicupdate}.
On the other hand, however, we find that many open source QUIC stacks do not mirror ECN~\cite{quicinterop}; moreover, a passive measurement study finds low ECN usage for suspected QUIC traffic on a university network~\cite{lim:arxiv22:freshlookecn}.
Aggravantingly, mirroring ECN is only half of what is required for ECN with QUIC: QUIC employs an ECN validation stage to check if the network path impairs ECN, leading to a possible deactivation of ECN with QUIC.
Hence, for ECN with QUIC in the Internet, it is currently unknown whether it can be used, and, if not, whether general support is missing or ECN validation fails.

In this work, we thus revisit the state of ECN on the Internet, changing focus from the TCP-centric view of previous studies to the QUIC-based web running HTTP/3.
Starting with an initial assessment of QUIC ECN mirroring, we deepen our study and analyze ECN validation, impairments on the network layer influencing ECN validation, and the influence of vantage point location on network impairments and, thus, ultimately, ECN with QUIC.

In general, our study is four-fold:
We first conduct large-scale measurements from our main vantage point on more than 180\,M domains to identify QUIC-capable websites and gather QUIC connections subject for ECN testing.
Secondly, we analyze these QUIC connections for ECN support and assess the ECN capability of the underlying stacks.
Thirdly, shedding light on ambiguous results, we employ tracebox-based~\cite{detal:imc13:tracebox} network tracing to analyze potential network impairments hindering QUIC ECN usage.
Finally, to circumvent potential impairments on the path from our main vantage point, we distribute our measurements to global vantage points to visit QUIC ECN support on a broader scale, verify our analysis, and attribute global ECN support.
\\
In detail, our contributions are as follows:

\vspace{-0.5em}
\begin{itemize}[noitemsep,topsep=5pt,leftmargin=9pt]
\item We present a distributed, large-scale methodology to assess ECN support with QUIC on more than 180\,M domains.
\item We observe a large majority of HTTP/3 webservers (\SI{80}{\percent} of hosts, \SI{94}{\percent} of domains) to not mirror ECN, thereby hindering ECN usage. However, we see a notable increase in ECN support over time.
\item We find QUIC's ECN validation to identify impairments invisible to TCP, further reducing ECN capability onto <\SI{0.5}{\percent} of domains and <\SI{2}{\percent} of webservers; Some of these impairments also impact ECN innovations such as L4S~\cite{RFC9330}.
\item We observe similar impairments everywhere around the world hindering ECN usage with QUIC globally.
\end{itemize}
\vspace{-0.5em}
\centerline{Our tools and our toplist data are available via \cite{sander:zenodo:dataset} and }
\centerline{\href{https://github.com/COMSYS/quic-ecn-measurements}{https://github.com/COMSYS/quic-ecn-measurements}}

\section{ECN with QUIC vs. ECN with TCP}
\label{sec:background}
\ac{ECN}~\cite{RFC3168} enables routers along an Internet path to add explicit congestion markings to forwarded packets instead of dropping them in times of congestion.
It thus allows to avoid inherent overheads of packet loss, such as retransmissions and their latency or FEC and their bandwidth overhead, as routers can directly inform the end-to-end congestion control of incipient congestion.
However, due to the in-band nature of ECN, markings are not received by the sender but by the receiver, which in turn needs to inform the sender of the signal.
Consequently, TCP and QUIC require support of the IP layer and of their remote transport endpoint.

\subsection{IP Layer Support}
On the IP layer, \ac{ECN} uses two bits of the former ToS field in IPv4 / two bits of the traffic class field in IPv6 for congestion information~\cite{RFC3168}.
These bits encode four codepoints to identify ECN Capable Transport (ECT) and mark packets for Congestion Experienced (CE):
\texttt{not-ECT\,(00)}\,if ECN is not supported,
\texttt{ECT(1)}\,and \texttt{ECT(0)\,(01\,and\,10)}\,if ECN is supported, and
\texttt{CE\,(11)} if congestion occurred.

\afblock{Router handling.}
If ECN is not supported, routers fall back to traditional packet drops.
Otherwise, they signal incipient congestion by setting the CE codepoint.
Originally, ECT(1) and ECT(0) had the same meaning to inform routers that traffic can react to ECN.
Yet, with the recent advent of L4S~\cite{RFC9330}, traffic marked with ECT(1) is handled differently on L4S routers and CE marks are applied more aggressively to give advanced congestion control a more precise view on queuing stats aiming to reduce queuing delays for low latency network services.

Independent of the exact CE semantics, however, markings only arrive at the receiver of the data stream due to the in-band nature of ECN on the IP layer.
They thus need to be mirrored, also requiring end-to-end support on the transport layer.

\subsection{Transport Layer Support}
\label{subsec:transport}
The receiver needs to mirror information on the received CE markings via the transport layer to enable suitable reactions by the sender.
Moreover, the transport layer must be able to cope with missing ECN support or impairments.
We begin by explaining how TCP tackled these issues, to then present the differences in how QUIC implements ECN.

\subsubsection{TCP}
\label{subsubsec:transport_tcp}
TCP allows mirroring received CE codepoints via its ECE (ECN Echo) flag~\cite{RFC3168}. %
Whenever a TCP segment with an active CE marking on the IP layer is received, TCP sends an ECE flag in its acknowledgment to inform the sender of the marking.
The sender then uses this signal for its congestion control and acknowledges the receipt via the CWR (Congestion Window Reduced) TCP flag to the receiver.

\afblock{Mutual, explicit negotiation.}
Yet, it is not guaranteed that every TCP endpoint is ECN-capable and can mirror the codepoints, even if recent updates of the TCP RFC~\cite{RFC9293} introduce ECN support as a "SHOULD" feature.
ECN support is thus mutually negotiated during the TCP connection handshake: %
the initial SYN packet is sent with active ECE and CWR flags and needs to be answered by a SYN-ACK packet with active ECE.
Only then mirroring of ECN is ensured and ECT may be set on outgoing IP packets.
While today's popular operating systems and their TCP stacks all support ECN~\cite{linuxipsysctl,windowsdctcp,bsdecn,macosecn}, it is not always enabled by default~\cite{windowsecn2008,systemddisablecn}, hindering the use of ECN with TCP.
Furthermore, the new web standard HTTP/3~\cite{RFC9114} switches from TCP to QUIC~\cite{RFC9000}. %

\subsubsection{QUIC}
\label{subsubsec:transport_quic}
In contrast to TCP, QUIC not only mirrors CE signals but all three ECN codepoints. %
For this, endpoints count the observed codepoints and report the counters via their ACK frames, providing very fine-grained information on the codepoints and supporting ECN extensions such as L4S.
Furthermore, QUIC differs from TCP in how ECN is negotiated.
It allows the unidirectional use of ECN without negotiation but validates the received values, hence presenting a more conservative ECN check.

\afblock{Specification.}
The implementation of ECN mirroring is stated as a "MUST" feature of QUIC~\cite{RFC9000}.
However, the standard weakens this mandate for ECN support.
Specifically, it "MUST" be implemented only if the ECN fields are accessible through UDP sockets, which most of today's operating systems support~\cite{quicwgecn,linuxudpecn}.
Furthermore, the standard describes the possibility that QUIC stacks deliberately do not implement ECN, which stands in contrast to the initial "MUST".
This ambiguity is also discussed in a reported erratum of the standard from February 2023~\cite{ecnerrata}.

\afblock{Unilateral ECN validation.}
To account for incompatible endpoints and routers along a network path, QUIC defines an \emph{ECN validation} phase. %
Specifically, each endpoint \emph{independently} decides on ECN usage (\ie~setting IP ECN codepoints) and validates its forward path, which is different to TCP's negotiation of mutual ECN support.
In particular, ECN validation is more conservative as it not only checks for support on the transport layer as TCP's ECN negotiation does but also checks for correct ECN traversal along the network path.
\begin{figure}
\centering
\includegraphics{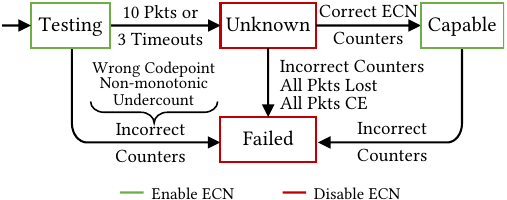}
\caption{Sample ECN validation algorithm of QUIC RFC 9000~\cite{RFC9000}}
\label{fig:quicvalidationalg}
\end{figure}
\pref{fig:quicvalidationalg} shows the validation algorithm in detail:
for the first few packets (the standard proposes to use 10), QUIC is in the testing phase and enables the ECT(0) codepoint to observe the mirroring results of the remote endpoint.
QUIC then disables the codepoints and waits for ACK frames from the remote side to assess ECN support.
If all packets are dropped, ECN validation fails and QUIC falls back to non-ECN mode.
Equally, missing/wrong ECN or overrepresented CE counters in ACKs lead to disabling ECN to account for routers zeroing ECN codepoints or marking every packet with CE.
If all ECN counters are correct, QUIC progresses into the ECN capable state and fully enables ECN usage.

\afblock{Terminology.}
We use the following terms to describe ECN support with QUIC:
\emph{Mirroring} if an endpoint mirrors ECN counters,
\emph{Capable} if the ECN validation succeeded,
\emph{Use} if an endpoint actively uses ECN (\ie~it sets ECN codepoints), and
\emph{Full Use} if ECN is used on an ECN capable path.

\section{Related Work}
\label{sec:relwork}
QUIC and TCP do not use ECN if it is not negotiated or if the validiation fails.
Hence, a widespread use of ECN is not guaranteed which is why many related works investigate the deployment and impairments of ECN on the Internet.

\paragraph{TCP ECN Support}
In 2001, Padhye and Floyd~\cite{padhye:ccr:tbit} analyzed the behavior of TCP stacks including the then new ECN feature. %
While they found only <\SI{1}{\percent} of \rsienvshort{24030} servers to be ECN capable in Sep. 2000, they observed a rapid adoption with $\sim$\SI{20}{\percent} of the same servers being ECN capable in Apr. 2001.
Medina et al.~\cite{medina:ccr:transportevolution} %
reran the ECN tests in 2004 for \rsienvshort{84394} servers, finding $\sim$\SI{2}{\percent} of servers to be ECN capable, where another $\sim$\SI{2}{\percent} did not correctly mirror ECN signals.
In 2008, Google~\cite{langley:techreport:tcpext} found for about \SI{10}{M} domains mapping to \SI{1.35}{M} active servers that around $\sim$\SI{1}{\percent} successfully negotiated ECN.

While all these works were limited in scope, or only tested specific ECN features, Bauer et al.~\cite{bauer:imc11:ecnreadiness} conducted extensive ECN measurements in 2011 on the Alexa Top 1M list.
They found $\sim$\SI{15}{\percent} of TCP-based webservers to be ECN capable, but most hosts in peer-to-peer networks ($\sim$\SI{95}{\percent}) to ignore ECN.
Using different vantage points, they further identified $\sim$\SI{17}{\percent} of paths to impair ECN, \ie~change or strip the ECT codepoints and ECN signals.

Extending these results, Kühlewind et al.~\cite{kuehlewind:pam13:ecnintheinternet} conducted ECN studies in 2012 via research and mobile networks.
For the Alexa Top 100k list, they found $\sim$\SI{27}{\percent} of webservers to support ECN.
While >\SI{90}{\percent} of those correctly mirrored ECN signals when contacted via the research network, no ECN feedback was observed in the mobile network which was attributed to performance proxies and middleboxes stripping ECN signals.
Additionally, the authors conducted the first measurements using IPv6, finding low IPv6 support (less than \SI{2.5}{k} hosts), but with higher ECN support ($\sim$\SI{48}{\percent}).
In later follow ups, the authors presented renewed and further findings:
starting off with \SI{56}{\percent} of IPv4 and \SI{65}{\percent} of IPv6 webservers supporting ECN in 2014~\cite{trammell:pam15:enablingecn}, adoption rose to \SI{74}{\percent} and \SI{95}{\percent} support for IPv4 and IPv6 in 2017~\cite{trammell:anrw17:pathspidertransportevol} and \SI{79}{\percent} for IPv4 in 2018~\cite{kuehlewind:tma18:ecnpathspider}.
However, the authors still found different kinds of ECN impairments on $\sim$\SI{1}{\percent} of the paths.

Mandalari et al.~\cite{mandalari:mcom18:ecnpp} found that TCP ECN extensions such as ECN++\cite{ecnpp} can be deployed with no further impairments in comparison to ECN.
For the Alexa Top 500k they find that \SI{61}{\percent} of servers support ECN via TCP.
Yet, they still find that the actual network path can impair ECN in many cases, especially in mobile networks where $\sim$\SI{65}{\percent} of the tested operators stripped ECN information.

The latest study published in 2022 by Lim et al.~\cite{lim:arxiv22:freshlookecn} presents measurements from mobile and wired networks on the Alexa Top 100k list, finding $\sim$\SI{86}{\percent} of hosts to negotiate ECN and \SI{4}{\percent} of paths to impair ECN signaling.
Studying passive measurements of a university network, they further find around \SI{5}{\percent} (port 80) to \SI{8}{\percent} (port 433) of webflows to mutually negotiate ECN and actually use ECN.

Overall, the presented studies observed a steady increase in TCP ECN support in the Internet from \SI{1}{\percent} in 2001 to around \SI{86}{\percent} in 2022. %
Yet, QUIC allows higher implementation flexibility being UDP-based and introduces a more conservative ECN validation phase which also tests the network path.
As such, prior TCP-based results cannot be mapped to QUIC.

\paragraph{ECN Support beyond TCP}

McQuistin et al.~\cite{mcquistin:imc15:ecnudp} presented the first work to shed light on the use of ECT codepoints with UDP in 2015. %
They found that $\sim$\SI{99}{\percent} of \SI{2.5}{k} NTP servers reachable via UDP were also reachable when using ECT, and that $\sim$\SI{1}{\percent} of network hops stripped ECT flags.
As such, using ECN with UDP was deemed possible, thus building the basis for ECN with QUIC.
Lim et al.'s~\cite{lim:arxiv22:freshlookecn} university network measurements also include UDP traffic where they identify $\sim$\SI{0.3}{\percent} of UDP port 443 flows (potentially QUIC) to set ECT codepoints in 2022.
However, they do not analyze the details of this low usage, if it really is QUIC traffic, and if ECN use or validation with QUIC are the limiting factors.

\takeaway{
To this date, ECN studies mostly focus on TCP, yet the results cannot be transferred to QUIC.
Only the works by McQuistin et al.~\cite{mcquistin:imc15:ecnudp} and Lim et al.~\cite{lim:arxiv22:freshlookecn} mention QUIC, but a large-scale view on ECN with QUIC on the modern web is still missing.
Moreover, while ECN support is formulated as a requirement of the QUIC standard~\cite{RFC9000}, most QUIC stacks on the QUIC interop runner~\cite{quicinterop} (11 out of 16) currently lack ECN support by design.
Joining this observation with the strong ECN support in TCP, it is unclear whether QUIC improves the web w.r.t. ECN support -- a gap we close with our study.
}

\section{Methodology}
\label{sec:method}
\begin{figure}
\centering
\includegraphics{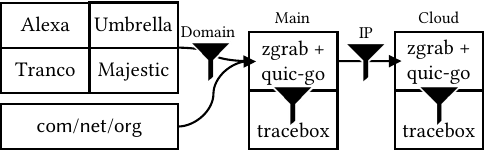}
\caption{Our measurements are fueled by \texttt{com}/\texttt{net}/\texttt{org} zone files and domain toplists. We measure these domains via zgrab and quic-go and start tracebox in case of irregularities. Moreover, we deduplicate our results and distributedly measure domains via cloud workers.}
\label{fig:methodpipeline}
\end{figure}

We perform large-scale measurements to assess the usage of ECN in the modern web.
First, we identify hosts of the web landscape for which we then assess ECN support.
Additionally, we analyze if any irregularities occur in the network.
Our overall measurement pipeline is shown in \pref{fig:methodpipeline}.
It is fueled by different domain lists, among which we use domain toplists compiled from the
(1) Alexa Top 1M\footnote{Please note that the Alexa Toplist is deprecated from May 2022 on and stopped changing in February 2023}~\cite{alexatoplist},
(2) Cisco Umbrella~\cite{ciscoumbrellatoplist},
(3) Majestic Million~\cite{majestictoplist}, and
(4) the Tranco Research List~\cite{pochat:ndss19:tranco}.
We update these toplists every Thursday, deduplicate their entries, and start our measurements on Friday.
In total, we gather around 2.7\,M unique domains from these lists.
Yet, these toplists are prone to change frequently~\cite{scheitle:imc18:topliststability} and do not map the broader web, such that we also use domain lists compiled from TLD zone files.
Here, we use the \texttt{com}/\texttt{net}/\texttt{org} zone files available at the ICANN Centralized Zone Data Service (CZDS)~\cite{icannczds}, which we download every Wednesday starting with our measurements.

On the identified hosts, we analyze the transport layer ECN support using HTTP requests via QUIC and TCP from our main vantage point in Aachen, Germany.
In particular, we validate ECN negotiation and mirroring, and evaluate ECN codepoint usage (§\ref{subsec:http_analysis}).
In case a host shows irregularities, we analyze the network layer connection to this host using a tracebox~\cite{detal:imc13:tracebox}-like approach similar to prior work on ECN support~\cite{lim:arxiv22:freshlookecn,mandalari:mcom18:ecnpp,kuehlewind:tma18:ecnpathspider,bauer:imc11:ecnreadiness,mcquistin:imc15:ecnudp}.
This allows us to identify possible impairments and/or changes to ECN codepoints while packets are in transit to further reason about missing ECN usage with QUIC (§\ref{subsec:network_analysis}).
Finally, we repeat the previous steps on distributed cloud instances to avoid router and middlebox influences on our path (§\ref{subsec:vps}).
Our tooling is available via \href{https://github.com/COMSYS/quic-ecn-measurements}{https://github.com/COMSYS/quic-ecn-measurements}.

\subsection{HTTP-based Transport Layer Analysis}
\label{subsec:http_analysis}
We issue QUIC-based HTTP/3 and TCP-based HTTP/2 / HTTP/1.1 requests to the identified hosts to deduce ECN support.
For this, we utilize zgrab2~\cite{zgrab2}, a golang-based banner grabber, which we adapt to additionally
(1) support HTTP/3, and
(2) log connection information.
For HTTP/3 support, we include a quic-go~\cite{quicgo} QUIC stack supporting QUIC v1 and extend it to also support the QUIC drafts 27, 29, 32, and 34 for long term measurements.
We adapt its retransmission behavior to only send 1 initial packet retransmission to reduce network traffic (\cf§\ref{sec:ethics}) and equip it with the ECN validation algorithm (\cf§\ref{sec:background}) as quic-go, by default, only mirrors ECN codepoints and does not use ECN on the forward path. %
Due to our adapted retransmission behavior, we reduce the validation phase to 5 packets and 2 timeouts.
Besides QUIC, we
(1) extend the TCP-based HTTP module to gather Linux's tcpinfo containing information about ECN negotiation success, and
(2) inject an eBPF program into the TCP socket to count ECN codepoints and log used TCP flags
in order to allow comparisons between QUIC and TCP.
Overall, our extensions allow us to observe in detail which ECN codepoints are used on the backward path and which signals have been mirrored for both QUIC and TCP to reason about ECN mirroring, validation, and usage.

We issue all requests with a \SI{10}{s} timeout to the "www" subdomain where we ignore redirects via Location headers or the Alt-Svc header to ensure comparability between QUIC and TCP and avoid repeatedly sampling the same domains. %
All requests are issued using HTTPS, and the IP address of the first DNS entry is always used for both QUIC and TCP.

\subsection{Tracebox-based Network Layer Analysis}
\label{subsec:network_analysis}
When the transport layer analysis shows abnormal behavior for a host, e.g., missing counters in QUIC or wrongly set ECN codepoints (e.g., ECT(1) instead of ECT(0)), we conduct a second measurement to analyze the network path to this host.
Similar to related work on ECN usage via TCP and UDP~\cite{lim:arxiv22:freshlookecn,mcquistin:imc15:ecnudp,mandalari:mcom18:ecnpp,kuehlewind:tma18:ecnpathspider,mcquistin:imc15:ecnudp}, we use a tracebox-based~\cite{detal:imc13:tracebox} approach sending QUIC Initial packets with active ECT codepoints and increasing TTLs to trigger ICMP time exceeded messages from the devices on path.
By varying the depth of path traversal through adjustments of the TTL, we can leverage the ICMP time exceeded messages containing a quotation of the original packet to shed light on the variations of the ECN codepoints during transit.
To account for lost ICMP packets, ICMP rate limiting, or blackholing, we leverage timeouts of \SI{3}{s} per hop, and accept up to 5 timeouts of subsequent hops before stopping the analysis.

\subsection{Distributed Cloud Instances}
\label{subsec:vps}
While we execute our main measurements from a single vantage point, we repeat both the transport and network layer measurements from distributed cloud instances to rule out router and/or middlebox interference on our path.
To reduce load and stress, and minimize the impact on networks and hosts (\cf§\ref{sec:ethics}), we deduplicate connections by IP and only forward the first viable domain per IP to the cloud instances if our main vantage point could successfully initiate an HTTP connection to the target host.
Hence, multiple requests to the same host (as issued by the main vantage point) are aggregated into a single request, which is then issued by each cloud instance.
This approach allows us
(1) to still deduce how many domains are reachable via QUIC, but
(2) deduplicate repeated requests to the same hosts (e.g., CDNs), and
(3) avoid connections to hosts blocking our main vantage point (\cf§\ref{sec:ethics}).
To account for geographic load balancing, each cloud instance locally resolves the forwarded domain name. %

\subsection{Limitations}
As with every measurement, our approach is subject to limitations restricting its findings and results.
For instance, we reduced the initial packet retransmission from 2 to 1 to reduce network stress (\cf§\ref{sec:ethics}) per domain.
I.e., our measurements may not establish connections in light of increased loss of the initial packets.
As part of this change, we also adapt the ECN validation phase to 5 packets and 2 timeouts.
This can influence our tests, especially in light of repeated CE signals that might be wrongly identified as all packets being marked with CE.
However, we see no signs of strong fluctuations of our measurements and for the "All CE" cases seen later on we see CE counts beyond 5 and also 10 packets.

Further, we refrain from using Alt-Svc headers and redirects for our measurement results to exactly argue about the domains that we shed light on and to avoid sampling domains that might not even be in our target set. %
Additionally, this decision allows us to directly compare QUIC and TCP throughout our evaluation, making sure both use the same host. %
Yet, this decision limits us in that we do not contact the intended QUIC server (for \rsienv{4595} domains / $\sim$\SI{0.02}{\percent} of all established QUIC connections) or are not redirected to the actual landing page of a website.

Another aspect is that we do not perform full-fledged webbrowser measurements, \ie~we do not request further resources that might be hosted at special content domains.
As such, we do not see all webflows, although we argue that many popular content domains are covered, e.g., through our usage of DNS-based toplists such as Cisco Umbrella.

For our tracebox-based network measurements, general limitations of traceroute apply.
Even though the measurements are conducted in parallel, the probes do not have to take the same routes as the zgrab measurements due to, e.g., load balancing or routing changes.

Finally, our decision to deduplicate website requests via their IP in our cloud-based measurements may introduce unwanted bias.
Our base assumption is that websites using the same IP are hosted at the same hoster and thus present equal behavior. %
This means that requesting the same website from another location may not direct us to the same server, but the same hoster with equal configurations.
While we believe this assumption to hold in many cases, approaches such as Meta-CDNs might induce a different behavior as websites are not necessarily hosted at the same hoster anymore.
Yet, we find this behavior for less than \SI{0.27}{\percent} of QUIC hosts. %
\begin{table}[t!]
\centering
\setlength{\tabcolsep}{0.4em}
\renewcommand{\arraystretch}{1.2}
\small
\begin{tabularx}{\columnwidth}{crrrXrrr}
\toprule
&
&
Total \# &
\multicolumn{1}{c}{Resolved \#} &
\multicolumn{1}{c}{QUIC \#} &&
\multicolumn{1}{c}{Mirroring} &
\multicolumn{1}{c}{Use} \\
\midrule
\multirow{2}{*}{\rotatebox{90}{\parbox{2.95em}{Toplists}}}  &
\#Domains &
\roundprefix[2]{2717571} &
\roundprefix[2]{1938214} &
\roundprefix[2]{525581} &&
\calcpercprecnosim[1]{17249}{525581} &
\calcpercprecnosim[1]{14747}{525581}\\
&
\#IPs &
&
\roundprefix[2]{785447} &
\roundprefix[2]{116701} &&
\calcpercprecnosim[1]{7878}{116701} &
\calcpercprecnosim[1]{6053}{116701}\\
\midrule

\multirow{2}{*}{\rotatebox{90}{\parbox{2.8em}{\centering c/n/o}}} &
\#Domains &
\roundprefix[2]{183278638} &
\roundprefix[2]{159398935} &
\roundprefix[2]{17304452} &&
\calcpercprecnosim[1]{970424}{17304452} &
\calcpercprecnosim[1]{718436}{17304452}\\
&
\#IPs &
&
\roundprefix[2]{9352326} &
\roundprefix[2]{232754} &&
\calcpercprecnosim[1]{45380}{232754} &
\calcpercprecnosim[1]{27569}{232754} \\
\bottomrule
\end{tabularx} 	\caption{Visible ECN Mirroring and Use via QUIC for toplist and com/net/org (c/n/o) domains using IPv4. Percentages are relative to the number of QUIC-capable domains and IPs.}
\label{tab:overviewquic}
\end{table}

\section{Visible QUIC ECN Support}
\label{sec:transport_layer}

Drilling down into QUIC ECN support in the web, we begin our assessment by analyzing the basis for ECN, i.e., whether ECN is mirrored and used, from our main vantage point.
In particular, we initially focus on the overall support of ECN with QUIC in the IPv4 space, evaluating which providers and domains occur in our latest measurements and identifying trends in our longitudinal data. %
Note that we do not yet consider the results of ECN validation, \ie~the ECN \emph{capability} (see §\ref{sec:validation}), or IPv6 (see §\ref{sec:ipv6}) at this point.

\subsection{Overview}
\label{sec:overviewquic}
\pref{tab:overviewquic} shows our IPv4 results (measured in week 15/2023) for the merged toplists and the \texttt{.com}, \texttt{.net}, and \texttt{.org} zone lists (c/n/o).
We denote how many domains and IPs were evaluated in total, how many could be resolved, and how many were QUIC-capable.
We further show the percentage of QUIC-capable domains and IPs that
(1) mirror ECN via QUIC, \ie~support the client in using ECN, and
(2) choose to use ECN themselves, \ie~set ECN codepoints on the IP layer (\cf§\ref{sec:background} for terminology).

\assert{\equal{\fpeval{7878/116701*100<7}}{1}}
\assert{\equal{\fpeval{7878/116701*100>6}}{1}}
\assert{\equal{\fpeval{6053/116701*100>5}}{1}}
\assert{\equal{\fpeval{6053/116701*100<6}}{1}}
\assert{\equal{\fpeval{45380/232754*100<20}}{1}}
\assert{\equal{\fpeval{45380/232754*100>19}}{1}}
\assert{\equal{\fpeval{27569/232754*100>11}}{1}}
\assert{\equal{\fpeval{27569/232754*100<12}}{1}}
\assert{\equal{\fpeval{17249/525581*100<4}}{1}}
\assert{\equal{\fpeval{17249/525581*100>3}}{1}}
\assert{\equal{\fpeval{14747/525581*100>2}}{1}}
\assert{\equal{\fpeval{14747/525581*100<3}}{1}}
\assert{\equal{\fpeval{970424/17304452*100<6}}{1}}
\assert{\equal{\fpeval{970424/17304452*100>5}}{1}}
\assert{\equal{\fpeval{802150/17304452*100>4}}{1}}
\assert{\equal{\fpeval{718436/17304452*100<5}}{1}}

We find that less than \SI{6}{\percent} of domains of the com/net/org zones and around \SI{3}{\percent} of the toplist domains
mirror ECN signals; thus, the QUIC standard's mandate of ECN mirroring is not widely visible.
Even fewer domains use ECN by setting ECN codepoints themselves, which is, however, not a requirement by the standard and can be chosen independently by the server.
Notably, the numbers significantly exceed the minor use of \SI{0.01}{\percent} found in related work~\cite{lim:arxiv22:freshlookecn}.
On a per IP-level, around \SI{20}{\percent} of hosts of the \texttt{com/net/org} domains, and around \SI{7}{\percent} of the \texttt{toplist} domains mirror ECN signals where the actual usage is again lower with around \SI{12}{\percent} and \SI{5}{\percent}.
In relative terms, we find that more hosts than domains support and use ECN, indicating that shared hosts, such as large-scale content providers serving a multitude of domains, might be hesitant to enable QUIC ECN mirroring and use, or are hidden behind ECN stripping network paths.
To avoid bias due to domain parking, we checked the \texttt{com/net/org domains} via NS/CNAME/A records~\cite{zirngibl:tma22:parking} where we identified \rsienv{108735} QUIC domains (\calcpercprecnosim[1]{108735}{17304452} of all QUIC \texttt{com/net/org} domains) to be related to domain parking; we therefore rule out a general bias in our data.

\takeaway{
We see ECN mirroring and usage for less than \SI{6}{\percent} of all QUIC-capable domains using IPv4.
In comparison, a higher share of IPs supports ECN mirroring, suggesting that large content providers might not support ECN via QUIC.
}

\begin{table}
\centering
\setlength{\tabcolsep}{0.45em}
\small
\begin{tabularx}{\columnwidth}{rrXlrrXrr}
\toprule

\multicolumn{1}{c}{$\uparrow$} & \multicolumn{1}{c}{Total \#} && AS Org. & Mirroring \# & \multicolumn{1}{c}{$\uparrow$} && Use \# & \multicolumn{1}{c}{$\uparrow$}\\
\midrule

1     &
\roundprefix[2]{8084195}   &&
Cloudflare   &
\roundprefix[2]{0} &
345 &&
\roundprefix[2]{0} &
255 \\

2     &
\roundprefix[2]{5652900}   &&
Google   &
\roundprefix[2]{145933} &
1 &&
\roundprefix[2]{0} &
255\\

3     &
\roundprefix[2]{1115011}   &&
Hostinger   &
\roundprefix[2]{111233} &
3 &&
\roundprefix[2]{81983} &
2\\

4     &
\roundprefix[2]{242595}   &&
Fastly   &
\roundprefix[2]{0} &
345 &&
\roundprefix[2]{0} &
255\\

5     &
\roundprefix[2]{152731}   &&
OVH SAS   &
\roundprefix[2]{49198} &
4 &&
\roundprefix[2]{38803} &
5\\

\midrule

6     &
\roundprefix[2]{137278}   &&
A2 Hosting   &
\roundprefix[2]{48996} &
5 &&
\roundprefix[2]{71265} &
3\\

7     &
\roundprefix[2]{128212}   &&
SingleHop   &
\roundprefix[2]{114424} &
2 &&
\roundprefix[2]{111857} &
1\\

8 &
\roundprefix[2]{87211} &&
Server Central &
\roundprefix[2]{0} &
345 &&
\roundprefix[2]{40436} &
4\\

\midrule

&
\roundprefix[2]{1704319} &&
<other> &
\roundprefix[2]{500640} &
&&
\roundprefix[2]{374092} &
\\

\bottomrule
\end{tabularx}
\caption{Top 5 providers of \texttt{com/net/org} QUIC-domains using IPv4 and whether they support or use ECN sorted by total domain count rank ($\uparrow$). Added are also the top 5 provider w.r.t. Use and Mirroring ($\uparrow$ describes the respective ranks).}
\label{tab:quicasnecn}
\end{table}

\subsection{Content Providers}
To analyze the impact of content providers on QUIC ECN usage, we next study the organizations in which our measured QUIC instances reside.
\pref{tab:quicasnecn} shows the AS organizations providing the websites of the \texttt{com/net/org} zones via QUIC and IPv4 sorted by the absolute number of domains provided / their rank ($\uparrow$).
Additionally, we also add the Top 5 ECN supporters by their ECN Mirroring and Use rank.
The AS organizations are inferred via CAIDA's as2org dataset~\cite{as2org} to fuse different ASs operated by the same provider\footnote{We also merge Cloudflare London and Cloudflare} together.
The respective ASs are mapped from our contacted IPs via BGP data of RIPE's RIS archive~\cite{riperis}.

\assert{\equal{Google}{Google}}

We find that large content providers such as Cloudflare and Fastly provide high shares of the overall domains supporting QUIC, but do not mirror ECN signals.
While Google ranks no. 1 in ECN mirroring in absolute numbers due to its massive scale, its relative support is rather low with \calcpercprecnosim[1]{145933}{5652900}.
Moreover, we do not find any ECN codepoints emitted by Google.
This overlaps with related work~\cite{lim:arxiv22:freshlookecn} which finds Google's AS to strip ECN information, although it contradicts the previous finding that ECN signals are mirrored by Google at all.
Medium sized providers such as Hostinger, A2 Hosting, or SingleHop show a much higher relative ECN mirroring support and also much higher ECN use in absolute numbers.
For example, Hostinger ranks no. 3 / 2 and contributes $\sim$110k / $\sim$80k domains with ECN mirroring / use.

When looking at the domain \texttt{toplists} (shown in \pref{tab:quicasnecntop}), we again observe that the medium sized providers make up large shares of the overall ECN support in relative numbers although their overall involvement in the toplists is much lower and thus explains the overall lower ECN support.
Moreover, Google's rank in ECN support is reduced significantly; we see that Google's own services (occurring often in the toplists, e.g., various per-country Google domains, YouTube/Google Video servers in the Umbrella list) do not mirror ECN via QUIC.
Instead, we see Amazon as the new no. 1 for ECN support and for ECN use.
We attribute this observation to the Amazon CloudFront CDN, which recently added HTTP/3 support~\cite{awscloudfronth3} and uses Amazon's own QUIC stack \texttt{s2n-quic} with known ECN support~\cite{quicinterop}.

\takeaway{
Large content providers mostly do not visibly support QUIC ECN mirroring or use on a large scale via IPv4.
Instead, adoption is mainly driven by smaller providers, resulting in low overall support.
Prominent exceptions are Google (only for \texttt{com/net/org}) and Amazon (only for \texttt{toplists}). %
}

\begin{table}
\centering
\setlength{\tabcolsep}{0.45em}
\small
\begin{tabularx}{\columnwidth}{rrXlrrXrr}
\toprule

\multicolumn{1}{c}{$\uparrow$} & \multicolumn{1}{c}{Total \#} && AS Org. & Mirroring \# & \multicolumn{1}{c}{$\uparrow$} && Use \# & \multicolumn{1}{c}{$\uparrow$}\\
\midrule

1     &
\roundprefix[2]{352477}   &&
Cloudflare   &
\roundprefix[2]{0} &
86 &&
\roundprefix[2]{0} &
72 \\

2     &
\roundprefix[2]{65916}   &&
Google   &
\roundprefix[2]{47} &
44 &&
\roundprefix[2]{0} &
72\\

3     &
\roundprefix[2]{12288}   &&
Fastly   &
\roundprefix[2]{0} &
86 &&
\roundprefix[2]{0} &
72\\

4     &
\roundprefix[2]{11642}   &&
Hostinger   &
\roundprefix[2]{1121} &
3 &&
\roundprefix[2]{830} &
5\\

5     &
\roundprefix[2]{3308}   &&
Amazon   &
\roundprefix[2]{3186} &
1 &&
\roundprefix[2]{3129} &
1\\

\midrule

7     &
\roundprefix[2]{2427}   &&
A2 Hosting   &
\roundprefix[2]{764} &
5 &&
\roundprefix[2]{1597} &
2\\

13     &
\roundprefix[2]{1455}   &&
SingleHop   &
\roundprefix[2]{1198} &
2 &&
\roundprefix[2]{1184} &
3\\

16 &
\roundprefix[2]{1129} &&
Interserver &
\roundprefix[2]{911} &
4 &&
\roundprefix[2]{853} &
4\\

\midrule

&
\roundprefix[2]{74939} &&
<other> &
\roundprefix[2]{10022} &
&&
\roundprefix[2]{7154} &
\\

\bottomrule
\end{tabularx}
\caption{Top providers of \texttt{toplist} QUIC-domains using IPv4 and whether they support or use ECN. (\cf\pref{tab:quicasnecn})}
\label{tab:quicasnecntop}
\end{table}

\subsection{Changes in ECN Mirroring over Time}
\label{subsec:changesecn}
Large providers are known to employ tests in production~\cite{cardwell:queue:bbr,singanamalla:imc22:origincf}.
Seeing the low support for ECN support by CDNs and top domains in our measurements in week 15/2023, we thus suspect that ECN support may have been enabled/disabled in the given week, potentially making our results a temporal artifact.
Hence, to assess the validity of our results, we next broaden our analysis to longitudinal data for the last year. %

\begin{figure}
\centering
\includegraphics{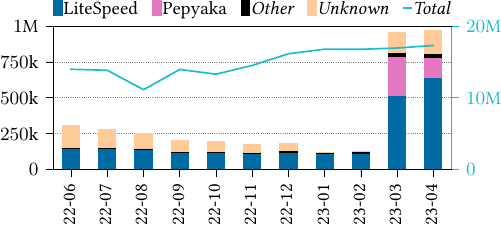}
\caption{HTTP/3 servers with observed ECN mirroring of \texttt{com/net/org} QUIC-domains over time using IPv4. Total (cyan line) refers to the right y-axis.}
\label{fig:quicservers}
\end{figure}

Figure \ref{fig:quicservers} shows the observed ECN mirroring of \texttt{com/net/org} QUIC-domains over time using IPv4.
We exclude ECN use for now in our results for clarity and being an optional feature.
The cyan line represents the total number of QUIC-domains and refers to the right y-axis; the stacked bars represent domains supporting ECN mirroring and refer to the left y-axis.
Despite excluding the \texttt{toplists} due to known instabilities (\cf§\ref{sec:method} and \cite{scheitle:imc18:topliststability}), we find a high variance over time:
For example, in Jun. 2022, \roundprefix[0]{\fpeval{13979762-13672652}} domains (\calcpercprec[2]{13979762-13672652}{13979762}) mirrored ECN via QUIC, while in Feb. 2023 only
\roundprefix[0]{\fpeval{16764518-16636151}} domains (\calcpercprec[2]{16764518-16636151}{16764518}) mirrored ECN.
In Mar. 2023, ECN mirroring considerably increased to \roundprefix[0]{\fpeval{17304452-16334028}} domains (\calcpercprec[2]{17304452-16334028}{17304452}).
In contrast, the overall support of QUIC is growing over time with little disturbances.

\begin{figure}
\includegraphics{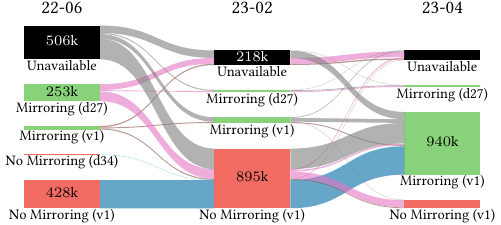}
\caption{Changes of QUIC ECN support over time with QUIC version (v) / draft (d) number in parentheses.}
\label{fig:quicsankey}
\end{figure}

\afblock{Impact of QUIC version.}
The sudden increase in ECN support could be rooted in
(1) domains previously unavailable via QUIC adding QUIC support and directly mirroring ECN, or
(2) available QUIC domains beginning to mirror ECN.
As such, we follow the temporal development of ECN mirroring per \texttt{com/net/org} domain in \pref{fig:quicsankey}.
We show how many domains have been unavailable, i.e., how many domains we could not contact via QUIC before, and how many available domains mirrored or did not mirror ECN.
Besides ECN mirroring, we also inspect the used QUIC version (e.g., \texttt{v1} for QUICv1, \texttt{d27} for QUIC draft 27).
Please note that we have filtered the plot for visibility by omitting changes involving less than \SI{3}{k} domains and removing paths that do not traverse through any form of ECN support. %
An unfiltered version of the plot can be found in the appendix (\pref{fig:quicsankey_nofilter}).

Looking at the results, the majority of connections with mirrored counters in Jun. 2022 (\roundprefix[0]{253141}) used the deprecated QUIC draft 27.
Most of these domains changed to QUIC version 1 without ECN support (\roundprefix[0]{106171}) or became unavailable (\roundprefix[0]{86763}).
Thus, in Feb. 2023, the ECN mirroring decreased to \calcpercprecnosim[2]{16764518-16636151}{16764518} while only few domains stayed with QUIC draft 27 and few newly added domains directly used ECN with QUICv1.
In Apr. 2023, most of the new ECN support was gained from domains already using QUICv1 but starting to mirror ECN (\rsienv{\fpeval{422982+306692}+108462}).
I.e., the sudden increase in ECN support came mostly from domains switching on mirroring and not from further domains directly using ECN, while the decrease came from domains updating to QUICv1 without enabled ECN support or going offline via QUIC.

\afblock{Impact of webserver.}
Digging deeper into the reasons for these changes in mirroring behavior, we next analyze the HTTP/3 server header (removing suffixes after \texttt{/} to ignore version numbers) returned by the contacted domains to identify the used webserver implementations and their QUIC stacks.
As can be seen in \pref{fig:quicservers}, the majority of webservers identified as \texttt{LiteSpeed}, followed by \texttt{Pepyaka} which is used by the website builder service wix.com and uses a Google reverse-proxy (the HTTP header via is set to "1.1 google").

To further identify the servers which did not set the server header (i.e., \texttt{unknown}), we compared the transport parameters of the QUIC connections and found that these were mostly equal to those of requests identifying as \texttt{LiteSpeed} and the LiteSpeed vendor website.
Hence, we deduce that most QUIC hosts in the wild mirroring ECN use \texttt{LiteSpeed}'s QUIC stack, where \texttt{LiteSpeed} also leads the charts for the \texttt{toplist} domains (not shown).
We also trace the transport parameters of the wix.com requests and find them to match Google's QUIC parameters.
In other words, Google's QUIC stack, when proxying wix.com requests, mirrored ECN signals in April while it did not for Google's own services.
This observation also explains the differences between the \texttt{toplist} and \texttt{com/net/org} domains for Google (\cf\pref{tab:quicasnecn} and \pref{tab:quicasnecntop}) as wix.com-hosted domains mostly feature in \texttt{com/net/org}.

\afblock{Pinpointing the causes.}
Overall, we identify incremental updates of server software, specifically \texttt{LiteSpeed} from QUIC draft 27 to QUICv1 and at Google's proxy, as likely reasons for the steady decline in ECN support until Feb. 2023 and its abrupt increase in Mar. 2023. %
However, changes on the network layer can also affect ECN mirroring, \eg{} introducing or resolving ECN impairments. %
Hence, the abrupt increase in ECN support might not solely be rooted in server software changes, but might also be grounded in changes of network configurations which we explore in §~\ref{sec:network}.

Nonetheless, for the \texttt{LiteSpeed} QUIC stack, we identified source code changes in Mar. 2023 w.r.t. accessing ECN on different platforms~\cite{litespeedrepochange,lsquicupdate} and an update for their freely available open source version~\cite{litespeedolsupdate}.
While these changes could explain the abrupt increase in ECN support we observed, we additionally contacted \texttt{LiteSpeed} as well as smaller providers (such as A2Hosting) about whether updates changed their ECN behavior, but did not receive an answer such that we cannot ascribe the specific root causes.
For Google, we also found git commits on their quiche QUIC stack which hint at ECN tests in Jan. and Mar. 2023~\cite{googleecn1,googleecn2}.

\takeaway{
We see significant variance in QUIC ECN mirroring over time.
The majority of webservers that provide domains with ECN support identify as \texttt{LiteSpeed} and version changes of its QUIC stack seem to enable / disable ECN support.
We also notice ECN experiments by Google in the last two months of our measurements.
Yet, changes in ECN support might not solely be rooted in changes of QUIC stacks, but might also be grounded in changes of network configuration which we explore in the next section.
}

\section{Clarifying Missing Support}
\label{sec:network}
In the first step of our analysis, we have focused on the server support for ECN with QUIC, finding little adoption -- a glaring difference to previous results on TCP (\cf§\ref{sec:relwork}).
However, network impairments of ECN traffic have a more drastic effect on QUIC compared to TCP as it uses a finer view on the actual ECN signaling with its validation approach.
Hence, as a second component of our root cause analysis for the small ECN support, we now turn our focus to the network layer, analyzing in detail whether ECN codepoints were impaired prior to reaching the destination.

\subsection{Cleared ECN Codepoints}
We start our analysis of possible network impacts with a tracebox-like network tracing approach which allows us to analyze whether ECN codepoints have been cleared on the client forward path, \ie~the path from the client to the server, as the reverse path remains hidden (\cf§\ref{subsec:network_analysis}).
Whenever we detect abnormal behavior (e.g., missing ECN mirroring) during our transport layer analysis (\cf§\ref{sec:transport_layer}), we analyze the corresponding network path.
For this, we send QUIC Initial packets with increasing TTLs along the path to trigger ICMP time exceeded messages, using the embedded ICMP packet quotation to identify potential codepoint clearing (see §~\ref{sec:method}).
Please note that the forward path of the tracing is not necessarily the same path as that of our transport layer analysis due to load balancing or route changes in-between.

\begin{table}
\centering
\setlength{\tabcolsep}{0.2em}
\small
\begin{tabularx}{\columnwidth}{lXrrXrrXrr}
\toprule

AS Org. && Cleared & \multicolumn{1}{c}{$\uparrow$} && Not Tested & \multicolumn{1}{c}{$\uparrow$} && Not Cleared  & \multicolumn{1}{c}{$\uparrow$} \\
\midrule

Server Central &
&
\roundprefix[2]{86948} &
1 &
&
\roundprefix[2]{263} &
27 &
&
\roundprefix[2]{0} &
439 \\

A2 Hosting &
&
\roundprefix[2]{78983} &
2 &
&
\roundprefix[2]{3470} &
4 &
&
\roundprefix[2]{5829} &
51 \\

Hostinger &
&
\roundprefix[2]{20049} &
3 &
&
\roundprefix[2]{20783} &
1 &
&
\roundprefix[2]{962946} &
3 \\

Contabo &
&
\roundprefix[2]{17253} &
4 &
&
\roundprefix[2]{741} &
10 &
&
\roundprefix[2]{930} &
159 \\

Sharktech &
&
\roundprefix[2]{16967} &
5 &
&
\roundprefix[2]{26} &
95 &
&
\roundprefix[2]{0} &
439 \\
\midrule

Fastly &
&
\roundprefix[2]{17} &
112 &
&
\roundprefix[2]{192} &
31 &
&
\roundprefix[2]{242386} &
4 \\

OVH SAS &
&
\roundprefix[2]{0} &
129 &
&
\roundprefix[2]{2101} &
5 &
&
\roundprefix[2]{101432} &
5 \\

Google &
&
\roundprefix[2]{0} &
129 &
&
\roundprefix[2]{11144} &
3 &
&
\roundprefix[2]{5495823} &
2 \\

Cloudflare &
&
\roundprefix[2]{0} &
129 &
&
\roundprefix[2]{11383} &
2 &
&
\roundprefix[2]{8072812} &
1 \\
\midrule

<other> &
&
\roundprefix[2]{110047} &
&
&
\roundprefix[2]{21905} &
&
&
\roundprefix[2]{1048904} &
\\
\midrule

<total> &
&
\roundprefix[2]{330264} &
&
&
\roundprefix[2]{72025} &
&
&
\roundprefix[2]{15931739} &
\\

<total IPs> &
&
(\roundprefix[2]{7704}) &
&
&
(\roundprefix[2]{50699}) &
&
&
(\roundprefix[2]{132587}) &
\\

\bottomrule
\end{tabularx}
\caption{Number of domains affected by detected ECN codepoint clearing per AS Organization using IPv4.}
\label{tab:traceboxclearingas}
\end{table}

\afblock{Result overview.}
\pref{tab:traceboxclearingas} presents the results for the \texttt{com/}\\\texttt{net/org} zones.
We split the results per AS and show the number of domains with and without visible ECN codepoint clearing (\emph{Cleared}/\emph{Not Cleared}) per AS together with total numbers in the last two rows.
To avoid ICMP rate limits and overwhelming routers due to the large number of cases, we sample our tracebox measurements: we trace every IP only once and with a probability of \SI{20}{\percent} .
Still, IPs occurring more often, such as CDN IPs serving thousands of domains, have a higher probability of being tested such that a large number of domains is covered. %
We indicate the number of domains that we did not test in a dedicated column.

Looking at our results, we observe that \roundprefix[2]{330264} domains were indeed affected by impairments on the path while \roundprefix[2]{15931739} domains without ECN mirroring were free from visible impairments.
In particular, these results include most domains of Google and Cloudflare, adding further evidence that their stacks do not mirror ECN.
Due to subsampling, we did not test the path for \roundprefix[2]{72025} domains, \ie~we cannot make definite statements regarding potential path impairments.
However, many of the affected domains again cover content providers for which we did not find ECN codepoint clearing in the performed tests.
Consequently, while we can conservatively state that \rsienv{330264+72025} domains (\calcpercprecnosim[1]{330264+72025}{17304452}) might have been subject to visible ECN impairments, we conjecture that the actual number of affected domains could be much closer to \roundprefix[2]{330264}.
In both cases, the overall share of ECN mirroring via QUIC would still only rise marginally from \calcpercprecnosim[1]{970424}{17304452} (\cf\pref{tab:overviewquic}) to at most \calcpercprecnosim[1]{970424+330264+72025}{17304452}.

\assert{\equal{A2 Hosting}{A2 Hosting}}
\assert{\equal{A2 Hosting}{A2 Hosting}}
\assert{\equal{Server Central}{Server Central}}
\assert{\equal{Server Central}{Server Central}}

\afblock{Impacted Providers.}
Inspecting the impacted providers, we find that \calcpercprecnosim[0]{86948}{87211} of domains hosted by Server Central and \calcpercprecnosim[0]{78983}{137278} of the domains hosted by A2~Hosting could not mirror ECN codepoints in Apr. 2023 as the fields were cleared on the path and thus invisible to the QUIC stacks serving the domains.
However, Server Central supported ECN on 58.30k domains in our measurements in Jun. 2022 and we noticed a route change from routes via Level3 to routes via Arelion/Telia Carrier in Dec. 2022.
Hence, in this case, we suspect that network elements skew our results, explaining part of the drop in ECN support identified in \pref{fig:quicservers}.

In contrast, the subsequent increase in ECN support does not seem to be rooted in network changes.
In particular, test traces in Oct. 2022 and Nov. 2022 revealed lower path clearing rates than in Apr. 2023.
Thus, an increase in ECN mirroring rooted in network changes should have been visible earlier than in Mar. 2023.
We thus attribute the change in adoption to QUIC stack updates and not the network.

\afblock{Network Provider Impact.}
Investigating the reasons for codepoint clearing, we find one network operator in particular to be involved:
\roundprefix[2]{325682} of domains (\calcpercprecnosim[1]{325682}{330264}) see a router of ASN 1299 / Arelion (formerly Telia Carrier) clearing codepoints, i.e., the configuration of these routers impacts the usage of ECN with QUIC. %
We contacted the operator but did not receive an answer, such that we cannot attribute the exact causes, so whether these changes were due to legacy routers rewriting the complete ToS field instead of the DSCP bits or whether deliberate changes were applied.

\takeaway{
\calcpercprecnosim[1]{330264+72025}{17304452} of domains are visibly affected by impairments on the forward network path where ECN codepoints are cleared.
We identify mainly one ISP to cause these impairments.
For the remaining domains we cannot pinpoint ECN impairments to be the main reason for missing ECN mirroring.
}

\begin{figure}
\includegraphics{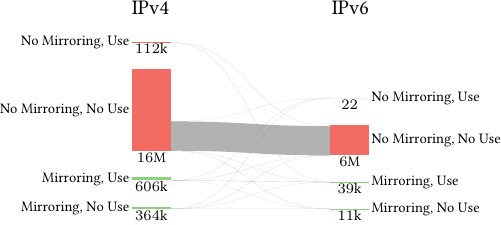}
\caption{IPv4 to IPv6 relation for visible ECN support.}
\label{fig:v4v6sankey}
\end{figure}

\subsection{IPv4 vs IPv6 Impact}
\label{sec:ipv6}
Seeing that routers are clearing ECN bits with IPv4, we set out to analyze whether these influences on ECN also occur with IPv6.
As such, we reran our measurements in week 13/2023 for IPv6 and compare them to our IPv4 measurements from week 15/2023.
Please note that IPv6 introduces completely new network layer and also new routes in comparison to IPv4.
Additionally, our subsequent measurements can introduce a temporal bias, e.g., due to route changes, for which we, however, found no indication in our data.

\pref{fig:v4v6sankey} shows how many domains of \texttt{com/net/org} mirrored / used ECN via IPv4 (left) and via IPv6 (right), further relating the behavior for domains present in both data sets. %
Overall, significantly fewer domains resolve to IPv6 addresses such that fewer QUIC domains can be seen.
Consequently, many of the QUIC domains reachable via IPv4 are unavailable via IPv6 which is particularly true for most of the domains supporting ECN with IPv4.
Indeed, 5\,M of 6\,M domains are served by Cloudflare without ECN mirroring and the high-rank ECN supporter A2 Hosting does not occur anymore (not shown).
Instead, we find that Google, Amazon, and Hostinger serve many of the QUIC-enabled and ECN-mirroring websites, but in very low absolute quantities (not shown).
As such, our IPv6 results are limited in their insights in comparison to IPv4 and we will present them only if noticeable differences occur.

Rerunning tracebox via IPv6 in week 16/2023, where we find similar zgrab results (not shown) to week 13/2023, we find 5 domains to be affected by codepoint clearing, while \rsienv{5550535} are not affected and \rsienv{646745} were not tested.
Combining these observations, we deduce that IPv6 routers impair ECN with QUIC less strongly, while the overall support still shrinks.

\takeaway{
Using IPv6, we find very little ECN clearing on the path impacting ECN mirroring.
Yet, barely any of the high-rank supporters of ECN via IPv4 provide their domains via IPv6.
As such, overall ECN support shrinks with IPv6.
}

\subsection{QUIC Mirroring vs TCP Mirroring}

The previous two sections indicate that network impairments might be a secondary cause for the low ECN support with QUIC.
However, our tracebox analysis is limited to the forward path, which can differ from prior measurements, and further relies on intermediate routers sending ICMP time exceeded messages.
Consequently, we cannot rule out path impairments completely.
Addressing this shortcoming, we next resort to a comparison between ECN support for TCP and QUIC: if domains/IPs correctly support ECN with TCP, then network issues (aside from middleboxes relying on information above layer 3) should not be the reason if the same domains/IPs do not support ECN with QUIC.

\afblock{Methodology.}
We repeat our IPv4 measurements in week 20/2023 for the \texttt{com/net/org} zones and access every website in parallel using TCP (HTTP/2 / HTTP/1.1) and QUIC (HTTP/3) while activating Linux's TCP ECN negotiation.
However, the negotiation alone does not allow us to analyze network layer influences as it is only an agreement to mirror CE codepoints, which might still be cleared by network devices.
Thus, we purposefully replace our ECT codepoints with CE codepoints (similar to related work on TCP~\cite{bauer:imc11:ecnreadiness}) to trigger TCP's ECE mirroring flag.
Equally, we set our QUIC ECN codepoints to CE to compare the results between QUIC and TCP.
Please note that this change can introduce different behavior on routers as CE instead of ECT(0) flags are seen; yet, this change is necessary to compare QUIC and TCP on equal footing.
As such, we examine these results in isolation from our previous findings.

\begin{figure}
\includegraphics{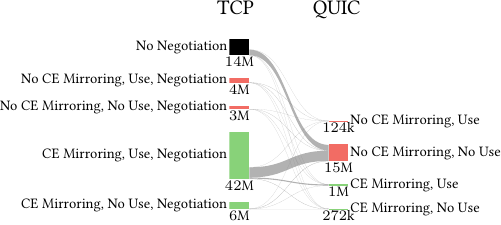}
\caption{TCP to QUIC Relation for visible ECN Support using IPv4.}
\label{fig:quictcpsankey}
\end{figure}

\afblock{Findings.}
\pref{fig:quictcpsankey} shows the results of our measurements, tracing for every domain whether it mirrored ECN via TCP (left) and QUIC (right) and connecting the respective bars accordingly.
\texttt{CE Mirroring} means that CE was correctly mirrored (via ECE for TCP / the corresponding counter for QUIC) while \texttt{Use} indicates that we saw ECN codepoints on the incoming IP packets.
Additionally, \texttt{Negotiation} means that ECN was negotiated between the TCP endpoints.

In our results, we observe that websites not supporting ECN via QUIC are mostly connected to two groups for TCP: one not negotiating ECN, the other negotiating, using and mirroring ECN correctly.
While the former group disallows assessing potential network impairments, the latter and larger group shows that successful use of ECN is possible for the domains.
I.e., this indicates that traditional network impairments do not limit ECN usage but that middleboxes or explicit decisions to not use ECN with QUIC hinder its usage.

\assert{\equal{Cloudflare}{Cloudflare}}
\assert{\equal{Google}{Google}}
\assert{\equal{Amazon}{Amazon}}

Inspecting the involved content providers in more detail (not shown), we again find that Cloudflare serves the most domains (\roundprefix[2]{9001571}) and supports ECN via TCP (use and mirroring) on \calcpercprecnosim[1]{24482+8975681}{9001571} of them.
Google serves \roundprefix[2]{8496389} domains and mirrors ECN without using it on \rsienv{1396902} while we cannot negotiate ECN for \roundprefix[2]{6532702} domains.
Amazon serves \roundprefix[2]{7861623} domains and supports ECN on \rsienv{164306+4903422}.
Thus, we see much higher ECN support with TCP overall.

Hence, the big QUIC providers are either obstructed by middleboxes that our prior tracebox measurements are unable to identify and that hinder ECN with QUIC in particular or they use QUIC stacks which do not support ECN.
Combining these results with the results of the QUIC interop runner~\cite{quicinterop}, we can specifically see that the opensource stacks of the providers and their deployed instances do not support ECN.

\takeaway{
The majority of domains that do not mirror CE codepoints via QUIC do mirror CE via TCP.
Of the remainder, the majority does not negotiate ECN via TCP and potentially disables ECN completely.
Barely any domains fail to mirror ECN via TCP while missing mirroring via QUIC, meaning that our tracebox analysis did not miss traditional impairments except for middlebox interference.
Yet, the results also correlate to the interop runner results of the dominant QUIC stacks.
As such, we deduce that the dominant QUIC stacks on the Internet seem to not mirror/use ECN in their current configurations.
}

\section{ECN Validation Challenges}
\label{sec:validation}
The previous two sections have assessed the general support for ECN with QUIC, indicated by mirroring and using ECN signals, which was mainly driven by QUIC stack support.
However, mirroring ECN is only one of two steps to \emph{fully use} ECN with QUIC as QUIC also validates the ECN counters:
the mirroring must match the codepoints that were actually sent.
Here, given QUIC stack support can still result in failed ECN validation due to network impacts beyond codepoint clearing.
In the following, we thus dive further into the mirrored ECN signals to assess whether ECN validation succeeds and whether ECN can be fully used on the forward-path.

\begin{table}
\centering
\setlength{\tabcolsep}{0.4em}
\small
\begin{tabularx}{\columnwidth}{lXrrXrr}
\toprule
&&\multicolumn{2}{c}{IPv4}&&\multicolumn{2}{c}{IPv6}\\
Mirrored Counters && IPs \# & Domains \# && IPs \# & Domains \# \\
\cmidrule{1-1} 	\cmidrule{3-4} 	\cmidrule{6-7}

All CE &&
\roundprefix[2]{2} &
\roundprefix[2]{4} &&
\roundprefix[2]{0} &
\roundprefix[2]{0} \\

Re-Marking ECT(1) &&
\roundprefix[2]{18548} &
\roundprefix[2]{301719} &&
\roundprefix[2]{6810} &
\roundprefix[2]{17147} \\

Undercount &&
\roundprefix[2]{22511} &
\roundprefix[2]{630579} &&
\roundprefix[2]{8388} &
\roundprefix[2]{27237} \\

Capable &&
\roundprefix[2]{4622} &
\roundprefix[2]{38116} &&
\roundprefix[2]{4956} &
\roundprefix[2]{5153} \\

\cmidrule{1-1} 	\cmidrule{3-4} 	\cmidrule{6-7}

No Mirroring &&
\roundprefix[2]{191112} &
\roundprefix[2]{16334028} &&
\roundprefix[2]{983489} &
\roundprefix[2]{6124271} \\

\bottomrule
\end{tabularx}
\caption{ECN validation results for the \texttt{com/net/org} domains.}
\label{tab:validation}
\end{table}

\subsection{Mirrored Codepoints and Validation}
\pref{tab:validation} shows the results of QUIC's ECN validation based on which ECN codepoints were mirrored for the \texttt{com/net/org} domains (\cf§\ref{subsubsec:transport_quic}).
For reference, we again show how many IPs and domains never mirrored ECN.

ECN validation succeeded (by progressing into the ECN Capable state) for \roundprefix[2]{38116} / \roundprefix[2]{5153} domains via IPv4 / IPv6 corresponding to only \calcperc{38116}{38116+6+301719+4+630579} / \calcperc{5153}{5153+0+17147+0+27237} of domains with ECN mirroring.
In these cases, QUIC's ECN validation does not detect issues on the forward path to the webserver and ECN can be used.
However, various issues occurred for the majority of domains with ECN mirroring, hindering the use of ECN.

For 4 domains retrieved via IPv4, ECN validation fails as all ECN codepoints were set to CE (we saw CE on more than 10 subsequent packets per connection).
Besides on-path routers illicitly marking all packets with the congestion codepoint, actual strong congestion is also a possible explanation of this effect.
With IPv6, we did not see this strong, repeated congestion marking.

For \roundprefix[2]{630579} / \roundprefix[2]{27237} domains (\calcperc{630579}{38116+6+301719+4+630579} / \calcperc{27237}{5153+0+17147+0+27237} of all domains which mirrored ECN), ECN validation failed due to signaling fewer ECN codepoints than were originally sent.
For instance, although 5 packets were sent with ECT(0) codepoints, only 3 packets were mirrored with ECT(0) set.

The remaining \roundprefix[2]{301719} / \roundprefix[2]{17147} domains (\calcperc{301719}{38116+6+301719+4+630579} / \calcperc{17147}{5153+0+17147+0+27237}) failed ECN validation due to mirroring ECT(1) instead of the originally sent ECT(0) codepoints.
We attribute this effect to two potential flaws:
(1) either the QUIC implementor mixed up ECT(0) and ECT(1), which can occur due to 2 (0b10) being ECT(0) and 1 (0b01) being ECT(1), or
(2) routers on the path re-mark packets from ECT(0) to ECT(1).

While the last observation does not affect the semantics of ECT according to the original ECN specification, and vanilla TCP does not even detect this change~\cite{RFC3168}, QUIC's ECN validation still reports it as a violation and disables ECN support.
Aggravatingly, network elements re-marking ECT(0) to ECT(1) can introduce issues with new applications such as L4S~\cite{RFC9330}.
Following the corresponding redefinition of ECT(1), ECT(1) now represents L4S support and causes routers to use a more aggressive marking method.
Consequently, network elements re-marking ECT(0) packets to ECT(1) on paths with L4S routers will mistakenly signal L4S support and potentially cause significantly reduced bandwidth for L4S-unaware congestion control.
In particular, traditional TCP implementations could suffer from serious performance penalties.

\takeaway{
ECN validation fails for the overwhelming share of QUIC endpoints that mirror ECN as ECT codepoints are signaled wrongly. %
Combined with the low general support for ECN mirroring, our client can only successfully use ECN with \calcpercprec[2]{38116}{17304452} /
\calcpercprec[2]{5153}{6173808} of all QUIC-capable domains.
}

\begin{table}
\centering
\setlength{\tabcolsep}{0.3em}
\small
\begin{tabularx}{\columnwidth}{lrXlrXlr}
\toprule
\multicolumn{2}{c}{Capable}&&\multicolumn{2}{c}{Undercount}&&\multicolumn{2}{c}{Re-Marking ECT(1)}\\
AS Org. & \multicolumn{1}{c}{\#} && AS Org. & \multicolumn{1}{c}{\#} && AS Org. & \multicolumn{1}{c}{\#} \\
\cmidrule{1-2} 	\cmidrule{4-5} \cmidrule{7-8}

Amazon &
\roundprefix[2]{19985} &
&
Google &
\roundprefix[2]{121424} &
&
A2 Hosting &
\roundprefix[2]{48991} \\

OVH SAS &
\roundprefix[2]{4693} &
&
SingleHop &
\roundprefix[2]{113343} &
&
Raiola Net. &
\roundprefix[2]{32384} \\

Hetzner &
\roundprefix[2]{2484} &
&
Hostinger &
\roundprefix[2]{79990} &
&
Hostinger &
\roundprefix[2]{31143} \\

PrivateSys. &
\roundprefix[2]{1525} &
&
OVH SAS &
\roundprefix[2]{44260} &
&
Google &
\roundprefix[2]{24483} \\

SingleHop &
\roundprefix[2]{1076} &
&
Interserver &
\roundprefix[2]{38574} &
&
Steadfast &
\roundprefix[2]{13272}  \\

\cmidrule{1-2} 	\cmidrule{4-5} \cmidrule{7-8}

<other> &
\roundprefix[2]{8353} &
&
<other> &
\roundprefix[2]{232976} &
&
<other> &
\roundprefix[2]{151446} \\

\bottomrule
\end{tabularx}
\caption{Number of domains affected by ECN validation results for the top-3 validation classes and the \texttt{com/net/org} domains per QUIC provider AS Organization using IPv4.}
\label{tab:validation_as}
\end{table}

\subsection{ASs and ECN Validation}
Seeing that only few connections can benefit from ECN due to a failed validation, we analyze the origin of the connections to assess which content providers are especially affected.
\pref{tab:validation_as} thus shows the involved AS organizations for the three biggest classes of ECN validation: successful validations, undercounted signals, and re-marking to ECT(1).

For the successful ECN validations, we see that mainly Amazon correctly mirrors signals such that primarily connections towards AWS can benefit from ECN with QUIC.
Conspicuously, our upstream provider (DFN - German Research Network) peers with Amazon in Frankfurt, i.e., only few routers lie between our vantage point and Amazon's endpoints.
We will analyze this observation further in §\ref{sub:remark_undercounting} and exploit it in §\ref{sec:global_view} for a global view on QUIC ECN.

On the flip side, the medium-sized content providers serving most of the domains mirroring ECN, do not pass ECN validation.
For these organizations, our upstream provider relies on transit providers.
The only exception to this correlation is Google, for which we either see little re-marking or strong undercounting despite a peering with our upstream.

\takeaway{
Amazon has got the largest share of domains that correctly mirror ECN signals and pass ECN validation.
In contrast, all previously identified medium-sized content providers and Google fail ECN validation due to ECT(1) remarking or undercounting.
We also identify a potential correlation between ECN validation and transit providers which we study next. %
}

\subsection{Clarifying Re-marking / Undercounting}
\label{sub:remark_undercounting}
To further analyze whether the QUIC stacks or the network paths re-marked ECT codepoints, we next analyze the path in cases of failed ECN validation using our tracebox measurements.
The results of these path traces are shown in \pref{tab:clarifying_remarking}.

\afblock{Undercounting.}
For undercounting of the ECN codepoints, which is the most significant class of ECN validation failures beside no mirroring at all, we barely find any routers along our path to mangle codepoints: \roundprefix[2]{629883} domains (\calcpercprecnosim[1]{629883}{630579}) show no changes along the path.
We thus attribute this observation to either intermittent mangling invisible to us or QUIC stack issues.
We looked into a few sample connections and found missing ECN information when \texttt{LiteSpeed} servers (which we identified for 4/5 of affected connections via transport parameters) switch between the handshake and 1-rtt packet number spaces.
We also reevaluated these sample connections with picoquic instead of our quic-go based pipeline to rule out a systematic issue with our quic stack and observed the same behavior.
Analyzing the open source code of \texttt{LiteSpeed}'s QUIC stack \texttt{lsquic}, we found that instances with a disabled ECN flag mirror ECN at first but do not transfer this setting to fully initialized connections.
Instances with enabled ECN flag mirror ECN in both cases.
The remaining connections originated from Google's AS (and used transport parameters which we previously found for Google's QUIC Stack), for which we also found re-marking issues.

\afblock{Re-marking.}
\begin{table}
\centering
\setlength{\tabcolsep}{0.2em}
\small
\begin{tabularx}{\columnwidth}{lXrrXrrXrr}
\toprule
&& \multicolumn{8}{c}{ECT(0) to} \\
&& \multicolumn{2}{c}{ECT(1)} && \multicolumn{2}{c}{Not-ECT} && \multicolumn{2}{c}{ECT(0)} \\
&& \multicolumn{1}{c}{IP} & \multicolumn{1}{c}{Domains} && \multicolumn{1}{c}{IP} & \multicolumn{1}{c}{Domains} && \multicolumn{1}{c}{IP} & \multicolumn{1}{c}{Domains} \\
\cmidrule{1-1} 	\cmidrule{3-4} \cmidrule{6-7} \cmidrule{9-10}

Re-Marking &&
\roundprefix[2]{16362} &
\roundprefix[2]{254745} &&
\roundprefix[2]{1935} &
\roundprefix[2]{22051} &&
\roundprefix[2]{251} &
\roundprefix[2]{24923} \\

Undercount &&
\roundprefix[2]{15} &
\roundprefix[2]{316} &&
\roundprefix[2]{3} &
\roundprefix[2]{135} &&
\roundprefix[2]{22261} &
\roundprefix[2]{629883} \\

\bottomrule
\end{tabularx}
\caption{ECN validation failures for the \texttt{com/net/org} domains and corresponding network impacts that could be seen for them.}
\label{tab:clarifying_remarking}
\end{table}
We find network elements to re-mark ECT codepoints to ECT(1) for \roundprefix[2]{254745} domains which affects the results for most of the smaller providers.
However, we also find \roundprefix[2]{24923} domains for which our tracebox approach cannot find any network impairments, indicating that the QUIC stack might incorrectly flag ECT(1).
We mainly observed this for Google's ASs; in conjunction with Google disabling ECN negotiation for Internet-facing TCP connections and based on related work assuming DCTCP usage in Google's ASs~\cite{lim:arxiv22:freshlookecn}, we suspect that Google might inadvertently expose its internal ECN signals via QUIC.

For \roundprefix[2]{22051} domains, we find codepoint zeroing although QUIC mirrors ECT(1), which we partly attribute to effects of load balancing / different routes on our tracebox analysis.

\assert{\equal{[[1299, 1299]]}{[[1299, 1299]]}}
\assert{\equal{[[2, 1]]}{[[2, 1]]}}

\assert{\equal{[[1299, 174]]}{[[1299, 174]]}}
\assert{\equal{[[2, 1]]}{[[2, 1]]}}

\assert{\equal{[[1299, 1299], [1299, 1299]]}{[[1299, 1299], [1299, 1299]]}}
\assert{\equal{[[2, 1], [1, 0]]}{[[2, 1], [1, 0]]}}

\afblock{Involved Network Providers:}
Analyzing where the routers changing the codepoints reside, we again find AS 1299 / Arelion / Telia Carrier for
\roundprefix[2]{162431} domains re-marking ECT(0) to ECT(1).
For \roundprefix[2]{92308} domains we see network elements residing in either AS 1299 (before) or AS 174 (Cogent, after visible change) re-mark our packets from ECT(0) to ECT(1).

We further observed elements of AS 1299 first re-marking packets from ECT(0) to ECT(1) and then from ECT(1) to not-ECT.
This occurred for \roundprefix[2]{16881} domains where our QUIC measurement reported re-marking to ECT(1).
We ascribe these variations to load-balancing at Telia / Arelion such that our trace took a different route than our transport layer measurement and hence traced slightly different router combinations at Telia.

\takeaway{
We cannot attribute ECN undercounting to network impairments, but instead suspect QUIC stack issues.
Indeed, we find that LiteSpeed's QUIC stack can undercount ECN explaining several cases.
Additionally, we find the same network operator as before to be involved in ECT re-marking, hindering ECN validation and, thus, the actual use of ECN with QUIC.
}

\section{Global View on QUIC ECN}
\label{sec:global_view}
The previous sections discuss the results of measurements conducted from a single vantage point, painting a picture that might be very specific to our location.
Hence, to evaluate QUIC ECN support on a global scale and to change/remove potential bias, we distribute our measurements to global locations of two cloud providers (AWS and Vultr) as the last piece of our study (\cf§\ref{sub:remark_undercounting}).

\begin{figure}
\includegraphics{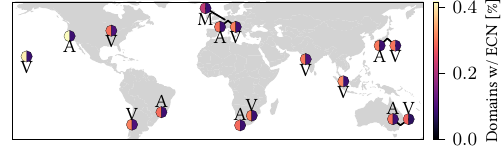}
\caption{Vantage points and domains that pass QUIC ECN Validation. \texttt{A} describes AWS, \texttt{V} Vultr, \texttt{M} main vantage point instances. Numbers scaled by domain to IP mapping of main vantage point. Left color represents IPv4, right IPv6 results.}
\label{fig:distributedvantagepoints}
\end{figure}

\pref{fig:distributedvantagepoints} shows our measurement vantage points and the respective ECN validation results.
The left half of each dot represents IPv4 support, while the right represents IPv6.
The measurements have been conducted in week 15/2023 (IPv4) and week 13/2023 (IPv6), where we deduplicate IPs at our main vantage point such that domains resolving to the same IP are requested only once via our cloud workers.
To still be able to argue about the number of domains involved, we rescale our results in this section by the previous domain-to-IP ratio.
Please note that in several vantage point locations requests to 3--9\,k IPs failed.
Converted back to domains, around 20\,k domains of the previously 17.30\,M QUIC-capable domains could not be requested, except for our most western instances in the US in Hawaii and San Francisco.
These instances were unable to build QUIC connections with 3\,k and 6\,k IPs some of which were heavy-hitters and mapped to in total 5\,M domains.
For both VMs, we found \rsienv{4953684} domains hosted via wix.com that switched from Google to wix.com infrastructure which did not support QUIC.
As such, results from these two measurement points need to be observed carefully.
We do not intersect the requests between the vantage points to avoid spreading a bias of one location to the others.
For the remaining vantage points, we find that for less than 70\,k mapped domains we are delegated to different ASs at our cloud vantage points, i.e., the impact is much smaller.

\afblock{Results.} Complementing our main vantage point results, we find a global QUIC ECN capability of around \SI{0.2}{\percent} to \SI{0.4}{\percent}.
Again, we find much more ECN mirroring for the domains of the \texttt{com/net/org} zones (around \SI{6}{\percent}, even up to \SI{20}{\percent} for India, not shown) than valid counter results.
This, once more, hints at network elements removing ECN information and thereby impairing ECN.
Support with IPv6 is again lower as observed before (\cf§\ref{sec:ipv6}).

\afblock{Location Bias.}
In general, we find the same ECN validation errors to occur again due to undercounting and re-marking of ECT(0) to ECT(1).
Yet, we see specific differences in the distribution of errors between some of the locations.
For example, 206 IPs from Google in India, mapping to \roundprefix[2]{23463} domains, always mirrored CE flags.
Also, we see an increased undercounting by Google in India by 516 IPs which map to  \roundprefix[2]{4975062} domains.
As we already saw varying numbers of ECN support from Google and recent changes on their QUIC stack w.r.t. ECN~\cite{googleecn1}, we conjecture that Google tested ECN more broadly in India.
Otherwise, we see very similar results as before with Hostinger, A2 Hosting, and Server Central being most affected by undercounting.
This issue does not seem to correlate with the location and its implicit changes to routing supporting our finding that this issue is not rooted in the network.

For re-marking, on the other hand, we see differences between the cloud providers.
For example, our Vultr instance in Central Europe (Frankfurt) sees less than 500 mapped domains with re-marking issues while AWS in Frankfurt sees more than 40\,k mapped domains to be affected.
Most re-marking is visible for our measurements in South America (Santiago de Chile) for nearly 100\,k domains.
This means that depending on the topological location, we can see strong changes in this ECN validation error category hinting again at network path issues.
Yet, these outliers barely affect the overall occurrence.
The total amount of network-induced errors stays even and comparable.

\afblock{Network Tracing.}
We thus also trace the network from our distributed vantage points to analyze which networks are involved.
Once more we often encounter AS 1299 / Telia Carrier / Arelion as the transit provider involved in re-marking packets.
For the majority of re-marking cases we find that at least one Telia router is involved.
Only in Central Europe (Frankfurt), Japan (Tokio), and the Central US (Chicago) our tracebox approach cannot identify a specific operator for the majority of cases.

\takeaway{
The global overall ECN support closely matches the percentages from our main vantage point, with barely more than \SI{0.3}{\percent} of domains passing ECN Validation.
Differences can be seen in the specific categories leading to failed ECN support where certain vantage points show particular spikes in contrast to others.
These spikes are, however, small enough such that they do not influence the overall picture.
Network-wise, again the same network operator as before seems to primarily account for packet re-marking affecting the validation process. %
Hence, the ability to use ECN with QUIC is currently also limited on a global scale.
} %
\section{Discussion}
Our work sheds light on the various aspects that are required for ECN to be used with QUIC and emphasizes that QUIC ECN support is vastly different from TCP's ECN support.
We see around \SI{70}{\percent} of domains mirroring ECN with TCP (\cf\pref{fig:quictcpsankey}), while not even \SI{10}{\percent} do so with QUIC.
Additionally, mirroring signals is only one half of successful ECN usage as QUIC's ECN validation also needs to be passed.
However, validation only succeeds for a minority of QUIC domains mirroring ECN with \calcpercprecnosim[2]{38116}{38116+6+301719+4+630579} for IPv4 and \calcpercprecnosim[2]{5153}{5153+0+17147+0+630579} for IPv6. %
In the following we discuss the extent of our results, impacts on standardization and signals not only for QUIC but also new, ECN-based techniques such as L4S~\cite{RFC9330}.

\subsection{Meaning of Results for QUIC}
With QUIC's focus on reducing latencies, we expected that ECN's ability to avoid costly retransmissions would be a strong incentive for its usage.
Therefore, we also expected higher ECN support than we ultimately measured.
Of course, QUIC is a rather young protocol and implementers might have focused on other parts of QUIC first leaving ECN for later milestones.
Yet, while the provider landscape for QUIC is dominated by large-scale content providers which can quickly adapt and adopt ECN, it is still fragmented.
Moreover, routers also need to correctly treat and mark packets for ECN avoiding packet loss.
Hence, we hypothesize that QUIC connections will still commonly encounter congestion loss in the (near) future.
Thus, we believe that research on the performance implications of congestion loss, such as its effects on HOL Blocking~\cite{sander:tma22:h3hol}, is still fruitful.

Additionally, however, we think that increased research on ECN with QUIC will help in spreading its deployment.
E.g., the performance implications of ECN with QUIC are not well-researched at the moment as is the actual ECN/AQM interplay on Internet paths.
A better understanding of both factors might offer incentives to both content and network providers to deploy ECN.

\subsection{Meaning of Results for Standardization}
The QUIC standard describes ECN mirroring as a "MUST" feature for implementations.
However, the specification is weakened in that ECN codepoints need to be accessible and also the further description of ECN support is vague.
For example, the specification also describes that ECN is unavailable if stacks decide to not implement it.
As part of this ambiguity, there already exists a reported erratum~\cite{ecnerrata} from February 2023.
The specification's ambiguity shows in the Interop Matrix of QUIC~\cite{quicinterop} and also in our results.

However, not all of the missing support of ECN seems to be rooted in ambiguity or implementers focusing on other aspects of QUIC.
For instance, we find several domains that neither negotiate ECN via TCP nor pass ECN validation:
these operators probably decided deliberately against ECN, even if recent updates of the TCP RFC~\cite{RFC9293} introduce ECN support as a "SHOULD" feature.
Hence, even a stronger, less ambiguously formulated "MUST" in the QUIC standard will probably not create \SI{100}{\percent} ECN support in the future.

\subsection{Meaning of Results for ECN in General}
Our results show that QUIC is often hindered from using ECN due to codepoint re-marking.
Traditional transport protocols such as TCP are unaware of these processes and continue using ECN normally.
QUIC, in contrast, presents a much more careful and conservative ECN validation method that meticulously checks ECN signals.
As such, it is debatable whether our finding merely shows that QUIC is too sensitive.
Yet, especially in the light of recent ECN innovations, we argue that it is actually a signal of potential ECN ossification that may hinder deployment of ECN innovations.

\afblock{Impact on ECN Innovations / L4S.}
While we did not observe any forms of ECN blackholing, the re-marking of ECT(0) to ECT(1) is a first step in ECN ossification where prior changes and decisions on header fields had no effect on transport protocols but do now.
For example, L4S routers may wrongly classify non-L4S traffic as L4S traffic due to re-marking.
This can result in L4S-managed queues filling up or frequent L4S signals disturbing TCP's classic congestion control, as briefly discussed in one of the L4S RFCs~\cite{RFC9331}.
Consequently, our results suggest that an impairment-free deployment of L4S (independent of QUIC) in the open Internet is hard to achieve at the moment.

\afblock{QUIC Incentives Against Re-Marking.}
Re-marking impairments go unnoticed by traditional TCP as it handles ECT(0) and ECT(1) equivalently, but pose an issue with ECN extensions and QUIC.
Yet, QUIC's approach to disable ECN in this case and the low ECN support in total will probably not incentivize solving these issues.

We thus recognize this as an opportunity to discuss whether it makes sense to also include ECN in QUIC's greasing approach (cf., spin bit greasing or QUIC bit greasing, i.e., enabling/disabling features randomly at a specific probability to avoid middleboxes ossifying on specific settings).
As such, we can imagine randomly enforcing a few ECN codepoints, for instance during the initial phase of a connection, to increase visibility of ECN even if ECN should not be used.

\section{Conclusion}
In this paper, we investigated whether it is possible to use ECN with QUIC in the wild.
As the basis for ECN usage with QUIC, QUIC endpoints are required to mirror ECN codepoints, yet we find \SI{80}{\percent} of hosts and \SI{94}{\percent} of domains providing websites via QUIC to \emph{not} mirror our ECN signals.
Additionally, even fewer hosts make use of ECN themselves.

The missing mirroring can be rooted in two aspects: (1) network impairments, or (2) QUIC stacks not supporting ECN.
Systematically tracing domains without mirroring, we find IPv4 routers mainly in one AS on the Internet to actually cause impairments; for IPv6, we barely find any. %
Thus, we primarily attribute the missing mirroring, which for many domains works via TCP, to the QUIC stacks.

However, even if ECN codepoints are mirrored, the signals are often invalid: in \SI{96}{\percent} of cases where ECN is mirrored, QUIC's ECN validation fails.
Hence, in total, only \SI{0.22}{\percent} of domains accessible via IPv4 and \SI{0.08}{\percent} accessible via IPv6 can actually make use of ECN on the forward path.
We pinpoint the main reasons to be (1) undercounted codepoints in the mirroring, and (2) routers remarking ECN codepoints from ECT(0) to ECT(1) which can again be traced to one ISP and can impact ECN innovations beyond QUIC.

Suspecting a potential bias from our vantage point, we finally distribute our measurements globally, yet still find results in the same range.
Overall, we thus conclude that using ECN with QUIC is currently severely limited.
\section*{Acknowledgments}
This work has been funded by the German Research Foundation DFG under Grant No. WE 2935/20-1 (LEGATO).
We thank Jörg Ott, Robin Marx, the anonymous reviewers, and our shepherd Stephen McQuistin for their valuable feedback.
We further thank the network operators at RWTH Aachen University, especially Jens Hektor and Bernd Kohler.

\bibliographystyle{ACM-Reference-Format}
\balance
\bibliography{reference}

\appendix
\section{Ethics}
\label{sec:ethics}
Our measurements do not involve users or their data.
To minimize the impact of our measurements we strictly follow widely accepted and agreed measurement guidelines and ethical measures of the internet research community~\cite{menloreport,durumeric:usenixsec12:zmap} and best practices of our institute.

\afblock{Load Reduction.}
We spread our measurements over multiple days to reduce load and stress on networks and end-hosts.
Additionally, our zgrab stack gathers several, distinct measurement results reused at our institute (e.g. in \cite{kunze:imc23:spinbitwild}) via one connection such that we avoid further measurements and, again, reduce load.
When distributing our measurements, we limit ourselves to very few measurements and also limit the extent.
E.g., our IP deduplication approach, while introducing a bias for, e.g., metacdn architectures, reduces load significantly (factor 40).

\afblock{Information and Opt-Out.}
Moreover, we inform about our research context, embedding our projectname as hint in every HTTP request and hosting websites on all measurement IPs explaining our study and how to be added to our blocklist to be excluded from further measurements.
Abuse E-mails are handled quickly and accordingly.

\afblock{Anticipating External Blocking.}
Our main measurements, causing most traffic, are conducted from within our network and are placed on a dedicated IP inside our measurement subnet.
Those providers that block our measurement subnet without an explicit blocklist opt out (we provide detailed subnet information, such that blocking is possible) are not contacted from our distributed vantage points leaving these operator measures intact.

\section{Appendix}

\begin{figure}[ht!]
\includegraphics{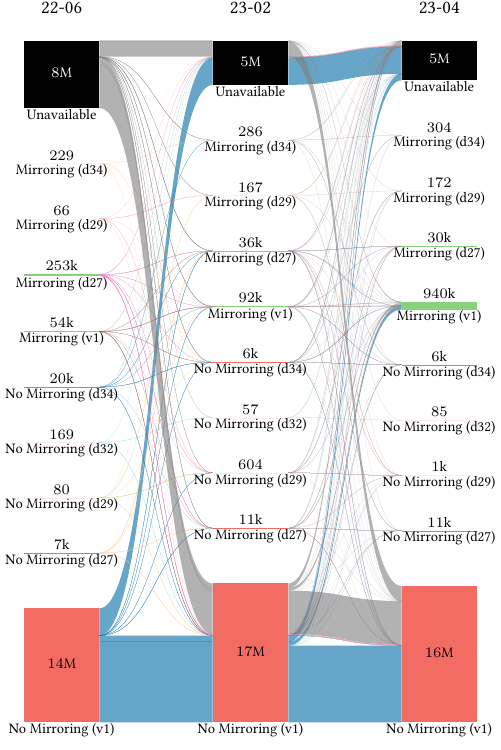}
\caption{Changes of QUIC ECN Support over time without filtering small changes}
\label{fig:quicsankey_nofilter}
\end{figure}

\end{document}